%% file: main.tex
\pdfoutput=1
\documentclass[journal]{IEEEtran}
\usepackage{hyperref}
\usepackage{framed,multirow}
\usepackage{cite}
\usepackage{amsmath, amssymb}
\usepackage{amsthm}
\usepackage{graphicx}
\usepackage{amsfonts}
\usepackage{longtable}
\usepackage{epsf, epsfig}
\usepackage{epstopdf}
\usepackage{subcaption}
\usepackage{cleveref}
\usepackage{cuted}
\usepackage{multirow}
\usepackage{algorithm}
\usepackage{algorithmic}
\usepackage{color}
\usepackage[normalem]{ulem}
\usepackage{booktabs}
\usepackage{adjustbox}
\usepackage{xspace}
\usepackage{doi}
\usepackage{xargs} 
\usepackage[pdftex,dvipsnames]{xcolor}



\graphicspath{{images/}}

\newcommand{\eat}[1]{}

\newcounter{gaocomm} 
  
\definecolor{blue-violet}{rgb}{0.54, 0.17, 0.89}
\definecolor{mygreen}{rgb}{0.0, 0.5, 0.0}
\definecolor{awesome}{rgb}{1.0, 0.13, 0.32}
\definecolor{bostonuniversityred}{rgb}{0.8, 0.0, 0.0}
\begin{document}

\title{Prognosis of Rotor Parts Fly-off Based on Cascade Classification and Online Prediction Ability Index}

\author{
%\thanks{Manuscript received xx x, 2021; revised xx x, 2021 and xx x, 2021; accepted xx xx, 2021. (Corresponding author: Jian Pu.)}
\thanks{Corresponding author: Zhe Song.}
\thanks{Y. J. Shen is with the Business School, Nanjing University, Nanjing, 210093, China (e-mail: dg1902054@smail.nju.edu.cn).}
\thanks{Z. Song is with the Business School, Nanjing University, Nanjing, 210093, China. (e-mail: zsong1@ nju.edu.cn).}
\thanks{A. Kusiak is with the Department of Industrial and Systems Engineering, 4627 Seamans Center for the Engineering Arts and Sciences, The University of Iowa, Iowa City, IA, USA (e-mail: andrew-kusiak@uiowa.edu).}

\normalsize{
Yingjun Shen,
Zhe Song\textsuperscript{$\ast$}, 
Andrew Kusiak
}
% <-this % stops a space
}

% \markboth{IEEE TRANSACTIONS ON NEURAL NETWORKS AND LEARNING SYSTEMS, VOL. XX, NO. XX, XXX 2020}%
% {Shell \MakeLowercase{\textit{et al.}}: Bare Demo of IEEEtran.cls for Journals}

\maketitle
\input{sections/abstract}


\input{sections/introduction}
\input{sections/problem_definition}

\input{sections/cascade_OPAI}

\input{sections/experiment}

\input{sections/conclusion}

\IEEEpeerreviewmaketitle

\bibliographystyle{IEEEtran}
\bibliography{biblio}

\begin{appendices}

\begin{table*}[]
\caption{The input features for N-A classifiers and description}
\centering
\label{table: N-A_inputs}
\begin{tabular}{@{}ccp{130mm}@{}}
\toprule
\textbf{Number} & \textbf{Features}  & \textbf{Description}                                                                                                    \\ \midrule
1               & Std\_S1            & Standard deviation of waves collected from sensor 1                                                                     \\
2               & FFT\_max\_S1       & Maximum amplitude based on FFT of waves collected from sensor 1                                                         \\
3               & Rms\_S1            & Root mean square of waves collected from sensor 1                                                                       \\
4               & Energy\_4\_por\_S1 & The proportion of the energy of the wavelet coefﬁcient of the fourth layer based on WT of waves collected from sensor 1 \\
5               & Energy\_3\_S4      & The energy of the wavelet coefﬁcient of the third layer based on WT of waves collected from sensor 4                    \\
6               & Energy\_3\_S1      & The energy of the wavelet coefﬁcient of the third layer based on WT of waves collected from sensor 1                    \\
7               & Max\_S1            & The maximum of waves collected from sensor 1                                                                            \\
8               & Energy\_3\_S2      & The energy of the wavelet coefﬁcient of the third layer based on WT of waves collected from sensor 2                    \\
9               & Peak\_Peak\_S1     & The peak-to-peak of waves collected from sensor 1                                                                       \\
10              & Energy\_2\_S5      & The energy of the wavelet coefﬁcient of the second layer based on WT of waves collected from sensor 5                   \\
11              & Rms\_S6            & Root mean square of waves collected from sensor 6                                                                       \\
12              & FFT\_std\_S1       & Standard deviation of amplitude based on FFT of waves collected from sensor 1                                           \\ \bottomrule
\end{tabular}
\end{table*}

\begin{table*}[]
\caption{The input features for R-H classifiers and description}
\centering
\label{table: R-H_inputs}
\begin{tabular}{@{}ccp{130mm}@{}}
\toprule
\textbf{Number} & \textbf{Features}  & \textbf{Description}                                                                                 \\ \midrule
1               & FFT\_max\_S2       & Maximum amplitude based on FFT of waves collected from sensor 2                                      \\
2               & FFT\_std\_S2       & Standard deviation of amplitude based on FFT of waves collected from sensor 2                        \\
3               & Std\_S2            & Standard deviation of waves collected from sensor 2                                                  \\
4               & Rms\_S2            & Root mean square of waves collected from sensor 2                                                    \\
5               & Energy\_3\_por\_S2 & The energy of the wavelet coefﬁcient of the third layer based on WT of waves collected from sensor 2 \\
6               & FFT\_max\_S1       & Maximum amplitude based on FFT of waves collected from sensor 1                                      \\
7               & Peak\_Peak\_S2     & The peak-to-peak of waves collected from sensor 2                                                    \\
8               & Min\_S2            & The minimum of waves collected from sensor 2                                                         \\
9               & Rms\_S1            & Root mean square of waves collected from sensor 1                                                    \\
10              & FFT\_std\_S1       & Standard deviation of amplitude based on FFT of waves collected from sensor 1                        \\
11              & Max\_S1            & The maximum of waves collected from sensor 1                                                         \\
12              & Std\_S1            & Standard deviation of waves collected from sensor 1                                                  \\ \bottomrule
\end{tabular}
\end{table*}

\begin{table*}[]
\caption{The input features for ternary classifiers and description}
\centering
\label{table: ternary_inputs}
\begin{tabular}{@{}ccp{130mm}@{}}
\toprule
\textbf{Number} & \textbf{Features}  & \textbf{Description}                                                                                                    \\ \midrule
1               & FFT\_max\_S1       & Maximum amplitude based on FFT of waves collected from sensor 1                                                         \\
2               & FFT\_std\_S1       & Standard deviation of amplitude based on FFT of waves collected from sensor 1                                           \\
3               & Rms\_S1            & Root mean square of waves collected from sensor 1                                                                       \\
4               & Std\_S1            & Standard deviation of waves collected from sensor 1                                                                     \\
5               & Energy\_3\_S4      & The energy of the wavelet coefﬁcient of the third layer based on WT of waves collected from sensor 4                    \\
6               & Energy\_3\_S2      & The energy of the wavelet coefﬁcient of the third layer based on WT of waves collected from sensor 2                    \\
7               & Peak\_Peak\_S1     & The peak-to-peak of waves collected from sensor 1                                                                       \\
8               & Energy\_4\_S1      & The energy of the wavelet coefﬁcient of the fourth layer based on WT of waves collected from sensor 1                   \\
9               & Max\_S1            & The maximum of waves collected from sensor 1                                                                            \\
10              & Energy\_3\_S1      & The energy of the wavelet coefﬁcient of the third layer based on WT of waves collected from sensor 1                    \\
11              & Energy\_4\_por\_S1 & The proportion of the energy of the wavelet coefﬁcient of the fourth layer based on WT of waves collected from sensor 1 \\
12              & FFT\_max\_S2       & Maximum amplitude based on FFT of waves collected from sensor 2                                                         \\
13              & Peak\_Peak\_S2     & The peak-to-peak of waves collected from sensor 2                                                                       \\
14              & Std\_S2            & Standard deviation of waves collected from sensor 2                                                                     \\ \bottomrule
\end{tabular}
\end{table*}

\begin{figure*}
\centering
\includegraphics[width=0.7\linewidth]{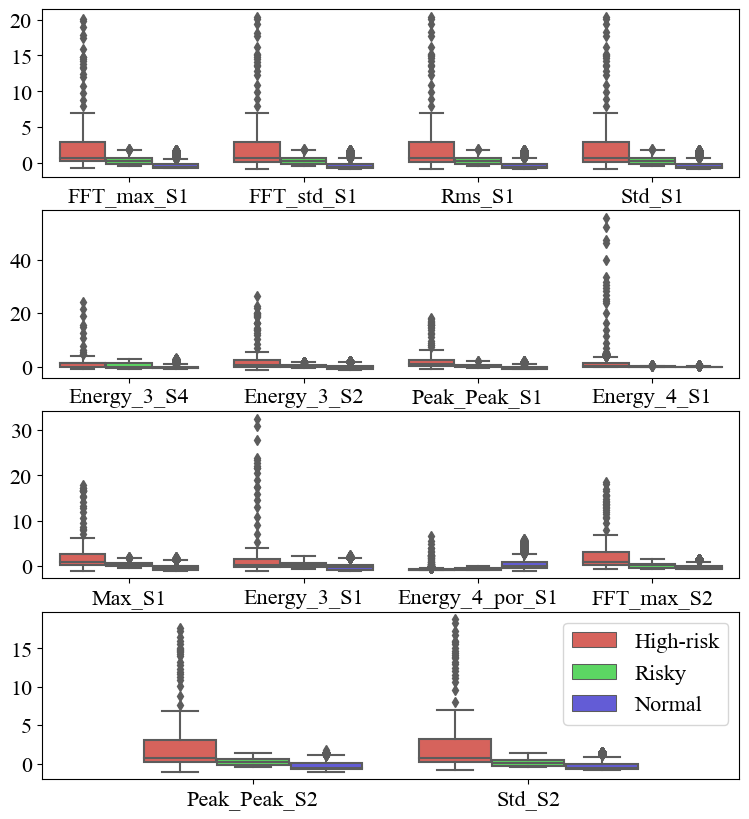}
\caption{Boxplots of input features for ternary classifying (left: High-risk, middle: Risky, right: Normal)}
\label{fig:ternary_features}
\end{figure*}

\end{appendices}

\end{document}

%% file: sections/abstract.tex
\begin{abstract}
Large rotating machines, e.g., compressors, steam turbines, gas turbines, are critical equipment in many process industries such as energy, chemical, and power generation. Due to high rotating speed and tremendous momentum of the rotor, the centrifugal force may lead to flying apart of the rotor parts, which brings a great threat to the operation safety. Early detection and prediction of potential failures could prevent the catastrophic plant downtime and economic loss. In this paper, we divide the operational states of a rotating machine into normal, risky, and high-risk ones based on the time to the moment of failure. Then a cascade classifying algorithm is proposed to predict the states in two steps, first we judge whether the machine is in normal or abnormal condition; for time periods which are predicted as abnormal we further classify them into risky or high-risk states. Moreover, traditional classification model evaluation metrics, such as confusion matrix, true-false accuracy, are static and neglect the online prediction dynamics and uneven wrong-prediction prices. An Online Prediction Ability Index (OPAI) is proposed to select prediction models with consistent online predictions and smaller close-to-downtime prediction errors. Real-world data sets and computational experiments are used to verify the effectiveness of proposed methods.
\end{abstract}

%% file: sections/introduction.tex
\section{Introduction}\label{Sec:1}
Rotor is the core part of large rotating machinery such as induction motor, compressor, steam turbine and gas turbine in aerospace, energy, chemical industries ~\cite{chen1995fault}. Due to the huge centrifugal force produced during the operation of large rotating units, rotors will become loose and fall off, which poses a great threat to safety. Rotor failures account for a large percentage of rotating machine failures, broken cage bars and bearing deterioration are the main causes of rotor failures ~\cite{benbouzid2003stator}. Condition monitoring and predictive maintenance have attracted higher attention for the demand of safer operation and less unscheduled downtime. However, rotor failure is usually a slow deterioration process, it is difficult to capture early symptoms.

Fault prediction methods include model-based approach and data-driven approach. Model-based approach uses physical models or domain knowledge to estimate the state of the system. Calculating the residual between the system’s observed value and model’s prediction, then, predicting the fault by comparing the residual with the threshold. Data-driven fault prediction methods include residual-based approach and supervised learning based predictive modeling. Residual-based approach mines historical normal data of the system to identify a data-driven model. Calculating the residuals between the online observed data and predictions of the identified model, when the residual exceeds the threshold, a fault is alarmed. Fault prediction based on supervised learning is relatively straightforward because the historical data set is analyzed by domain experts and each record is labeled with normal or abnormal information.

Traditional fault prediction methods, such as physical modeling ~\cite{simani2003model, isermann2005model}, causal analysis ~\cite{ekanayake2019model}, expert system ~\cite{gelgele1998expert}, signal processing and analysis ~\cite{frank2000model}, rely on failure mechanism deeply. Physical modeling predicts fault based on the deviation between observation and the estimator. The causal analysis uses symbol graph or fault tree to predict fault. Expert system generates a series of rules to predict faults by mimicking expert reasoning process. Signal processing and analysis includes time-domain, frequency-domain and time-frequency methods ~\cite{riaz2017vibration}. It is the most common and effective approach which predicts fault by analyzing, transforming and decomposing the collected time series signals, then comparing the changes of signal features such as magnitude, phase shift or frequency ~\cite{ding2008model, mokhtari2017vibration}. Time-domain methods include correlation analysis and signal model identification. Frequency-domain methods include Fourier Transform and cepstrum techniques. Time-frequency methods include wavelet transform, short-time Fourier transform ~\cite{kankar2011fault, wu2009expert}. Da Costa et al. (2015) compared the performance of Fast Fourier Transform (FFT) and Wavelet Transform (WT) methods in rotor fault prediction of asynchronous cage motors. They found that although FFT method is simple and robust, but not applicable in non-stationary conditions. WT can be used as supplement to FFT method in non-stationary condition ~\cite{da2015rotor}.

In-depth analysis of fault mechanism provides high accuracy, interpretability, reliability and generalizability to traditional methods. Physics-based fault prediction model has less parameters and higher interpretability. For example, different faults, such as winding, torque fluctuation, air gap eccentricity, will lead to different symptoms of parameters, such as vibration, current and acoustics ~\cite{gundewar2021condition}. However, traditional physics-based methods are difficult to predict faults in advance, e.g., abrupt, intermittent and incipient faults. Furthermore, strict boundary conditions make it hard to adapt to nonlinear and time-varying nature of complex systems. Various condition changes, such as season, temperature, irradiance or humidity changes ~\cite{tahan2017performance, vichare2006prognostics}, will cause misjudgment. Lastly, physics-based fault prediction model usually requires limited number of observable variables. However, some data are difficult to collect because of sensor problems in practice. On the other hand, additional available useful information cannot be included with regret. The noise and measurement deviation in data collection challenges the traditional fault prediction methods. With the development of digital factories, Supervisory Control and Data Acquisition (SCADA)) system collects massive data, but the traditional methods cannot make full use of Big Data. Data-driven fault prediction methods, which can handle a large amount of data, is increasingly popular ~\cite{xu2017industrial}.

Residual-based fault prediction essentially involves calculating the residuals between the system’s observed value and model’s prediction, and estimating the distribution of residuals, then, predicting the faults based on 6-sigma principle or residual plots ~\cite{oh2020residual, gnetchejo2021faults}. Kusiak and Song (2009) used a decision tree to model a boiler combustion process, and observe the prediction residuals’ distributions of the training and test sets. Different types of sensor faults could be identified based on 6-sigma control charts ~\cite{kusiak2009sensor}. Zhang et al. (2012) used a data mining algorithm to analyze the vibration data of a wind turbine gearbox and calculated the upper and lower bounds of the exponentially weighted moving average chart (EWMA) based on the residual distribution. Then, the EWMA chart is used to monitor the high-speed operations of the gearbox ~\cite{zhang2012fault}. The residual-based approach is suitable for scenarios where there are no physical models or fault labels (or it is expensive to obtain fault labels). For example, Hamadache et al. (2019) used a residual-based approach to predict the fault of railroad electromechanical switching system and found that residual-based approach is computationally fast and no need for pre-annotated labels, which can reduce the cost of obtaining fault labels ~\cite{hamadache2019residual}. Serdio et al. (2014) used a residual-based approach for mill fault prediction, by comparing a residual-based approach with traditional PCA, time series analysis, and support vector machine (SVM), they found that the residual-based approach has higher accuracy, shorter computational time and lower false-alarm rate ~\cite{serdio2014residual}.

Intelligent manufacturing and Industrial 4.0 have accelerated the adoption of intelligent sensors. As a result, manufacturing is becoming smarter and more digital. Intelligent sensing devices can collect large amounts of labeled data, which provide basis for supervised machine learning-based fault prediction. For example, different algorithms, Classification and regression trees (CART), k-nearest neighbor (KNN), Bayesian inference, support vector machine (SVM), and artificial neural network (ANN) have been used for fault prediction ~\cite{sun2007decision, cai2016model, hongm2009data, liu2018artificial}. Verma and Kusiak (2012) used 24 data mining algorithms to analyze the wind turbine SCADA data, and could predict wind turbine brush failure 12 hours in advance ~\cite{verma2012fault}. 

There is a trend to combine physical principles and machine learning algorithms in fault prediction, where important features are extract based on physical principles, then fault prediction models are built with popular machine learning algorithms. Freeman et al. (2021) used continuous Morlet wavelet transform to obtain waveform features, then a simple KNN algorithm is used to construct a turbine rotor blade imbalance fault prediction model, which achieved 100\% accuracy on a simulated dataset ~\cite{freeman2021rotor}. 

Due to the excellent performance in automatic feature extraction and self-learning, deep learning algorithms are increasingly used in fault prediction  ~\cite{cheng2017fault, zhang2019bp}. Deutsch and He (2017) predicted the remaining lifetime of rotating equipment based on the deep belief network-feedforward neural network (DBN-FNN), which proved that DBN-FNN was able to automatically extract features with high prediction accuracy ~\cite{deutsch2017using}. Kusiak and Verma (2012) used deep neural networks (DNN) to predict the wind turbine generator bearing fault, and validated on different data sets with an accuracy more than 97\% ~\cite{kusiak2012analyzing}. 

Another trend of fault prediction is hybrid method. Wang et al. (2016) used DNN and EWMA control chart for gearbox fault prediction, which predicted impending failures 2 to 3 days in advance ~\cite{wang2016wind}. Wen et al. (2017) converted signal data into images and extracted image features, then LeNet-5 based convolutional neural network (CNN) is used for prediction modeling. This method was tested on motor bearing dataset, self-priming centrifugal pump dataset and axial piston hydraulic pump dataset, and achieved higher prediction accuracy than traditional machine learning algorithms (e.g., adaptive deep CNN, DNN, SVM) ~\cite{wen2017new}. 

Supervised machine learning-based approach requires a large amount of labeled data, but obtaining labels is time-consuming and sometime expensive. Some scholars have tried to extract features based on unsupervised machine learning algorithms to improve the adaptive capability of data-driven fault prediction. Lei et al. (2016) used sparse filtering to extract features directly from vibration signals, then, SoftMax regression is used with the learnt features, which was validated on motor bearing and locomotive bearing datasets with high prediction accuracy ~\cite{lei2016intelligent}; Li et al. (2020) used an unsupervised model based on sparse self-coding, deep confidence networks, and binary processors for fault prediction, which was validated on bearing fault and gear pitting fault datasets ~\cite{li2020unsupervised}.

Comparing with traditional fault prediction methods, there are two advantages of data-driven methods. Firstly, they can take advantage of big data. Machine learning algorithms can handle high-dimensional features, especially with industrial big data with high dimensionality and data volume ~\cite{kuo2019data}. Secondly, they can predict faults earlier and faster. Machine learning algorithms, such as multilayer perceptron (MP) and CNN, have proved to achieve high prediction accuracy in many applications. 

However, existing data-driven fault prediction methods have limitations in two aspects. One is that the prediction model’s generalization ability is weak as the industrial condition is complex and changeable, e.g., different machine types, environmental condition and system parameters challenge the generalization ability of a data-driven model. The other is the model’s self-adaptive ability is weak. Changes of industrial condition require the model to have the self-renewal ability by utilizing new data, which is generally not reflected in current data-driven fault prediction model.

This paper will focus on two under-researched aspects of data-driven fault prediction methods. First, most studies neglected the deterioration process and multiple severity levels of fault. They usually divide training data points into two categories, and perform normal-abnormal binary classification. For example, bearing deterioration changes exponentially, the vibration magnitude is small at the beginning and difficult to detect. When the fault deteriorates into a high-risk stage, a sudden fracture may occur. Some research literatures attempted to predict the fault types, magnitude, as well as severity levels of fault. They are essentially a supervised learning problem with multiple fault labels. A classifier is built to predict the labels, which is complex and costly ~\cite{cerrada2018review}. Bordoloi and Tiwari (2014) used a SVM to predict four different types of gearbox faults, and optimized the SVM parameters by grid search, genetic algorithm (GA) and artificial bee colony algorithm (ABCA) ~\cite{bordoloi2014optimum}. Although the training error is small, the model is too complex with long computational time, which makes it difficult to meet the real-time prediction and low-computing cost requirements. The internal structures of a large industrial equipment are complex. Similar fault symptoms are caused by different types of faults. Thus accurate predictions of fault type, magnitude, and degree of deterioration are necessary for solving the industrial problems.

Second, existing model evaluation metrics are static without considering the time sequence training and testing data points. For example, commonly-used metrics for prediction accuracy: precision, recall, sensitivity, and specificity ~\cite{sachs2012applied}, or mean absolute error (MAE), mean absolute percentage error (MAPE), root mean absolute square error (RMSE), and percent mean absolute relative error (PMARE) used together with k-fold cross-validation ~\cite{ali2014new}, all are static. These metrics usually assign the same weight to each data point, and evaluate the performance of the prediction model by calculating the error between the predicted value and the actual value on the train or test sets. The prediction accuracy of each sample in different positions of a time series has the same impact on the model’s total accuracy, which can’t meet the actual needs of fault prediction. 

In practice, the closer to break down, the more dangerous the equipment is. Different operating states, such as normal, early deterioration, risky or high-risk mean different O\&M actions. Therefore, data points at different time sequences of deterioration process should have different scoring weights. As the deterioration degree increases, the weights should be increased either. 

Furthermore, most prediction model’s evaluation metrics are offline, e.g., numeric indices, graphical evaluation methods, including error rate, ROC curve, Precision-recall curve, cost curve, etc. These metrics evaluate the model’s performance based on offline historical data points. How to continuously monitor, evaluate the accuracy and stability of the models after they are deployed online is not well considered.

To address the above under-researched aspects, this paper proposes a new approach in rotor fault prediction, including the prediction modeling process innovation and a new evaluation index construction. 

In terms of prediction modeling process innovation, this paper proposes a cascade prediction algorithm. Classical fault prediction models determine whether the equipment is normal at first; if not, discover early faults and causes; finally, predict the fault deterioration trend ~\cite{liu2018artificial}. The proposed cascade prediction model uses a normal-abnormal classifier for prediction. If the machine is normal, no further computation is needed for deterioration prediction. Once the machine is diagnosed to be abnormal, a risky and high-risk classifier is applied. The proposed prediction process can identify early deterioration and perform continuous online monitoring. Thus, once a fault is detected, it can quickly alert O\&M personnel to implement actions, such as isolation and accommodation ~\cite{ekanayake2019model}. Furthermore, the integration of physical principles in machine learning process makes the deterioration prediction simpler and the prediction model more interpretable and generalizable. 

In terms of evaluation index innovation, Online Prediction Ability Index (OPAI) is proposed to evaluate the accuracy and robustness of prediction. Based on the time to the moment of failure, different weights are assigned to the data points to evaluate the prediction accuracy. In terms of online monitoring, an indicator is constructed to evaluate the robustness of the model. The effectiveness of the algorithm and evaluation indicators proposed in this paper are verified with the real-world data.

The remaining part of the paper is organized as follows. Section~\ref{Sec:2} briefly introduce the problem definition of rotor parts fly-off prediction. Section~\ref{Sec:3} introduce the Cascade classification and Online Prediction Ability Index (OPAI). Then, in Section~\ref{Sec:4}, computational experiments are proposed to verify the cascade classifying algorithm and OPAI. The process of data acquisition, preprocess, feature engineering and cascade classifying models building are introduced in detail. In Section~\ref{Sec:5}, we will discuss and conclude this paper.

%% file: sections/problem_definition.tex
\section{The problem definition of rotor parts fly-off prediction}\label{Sec:2}

\begin{figure}[h]
\centering
\includegraphics[width=1\linewidth]{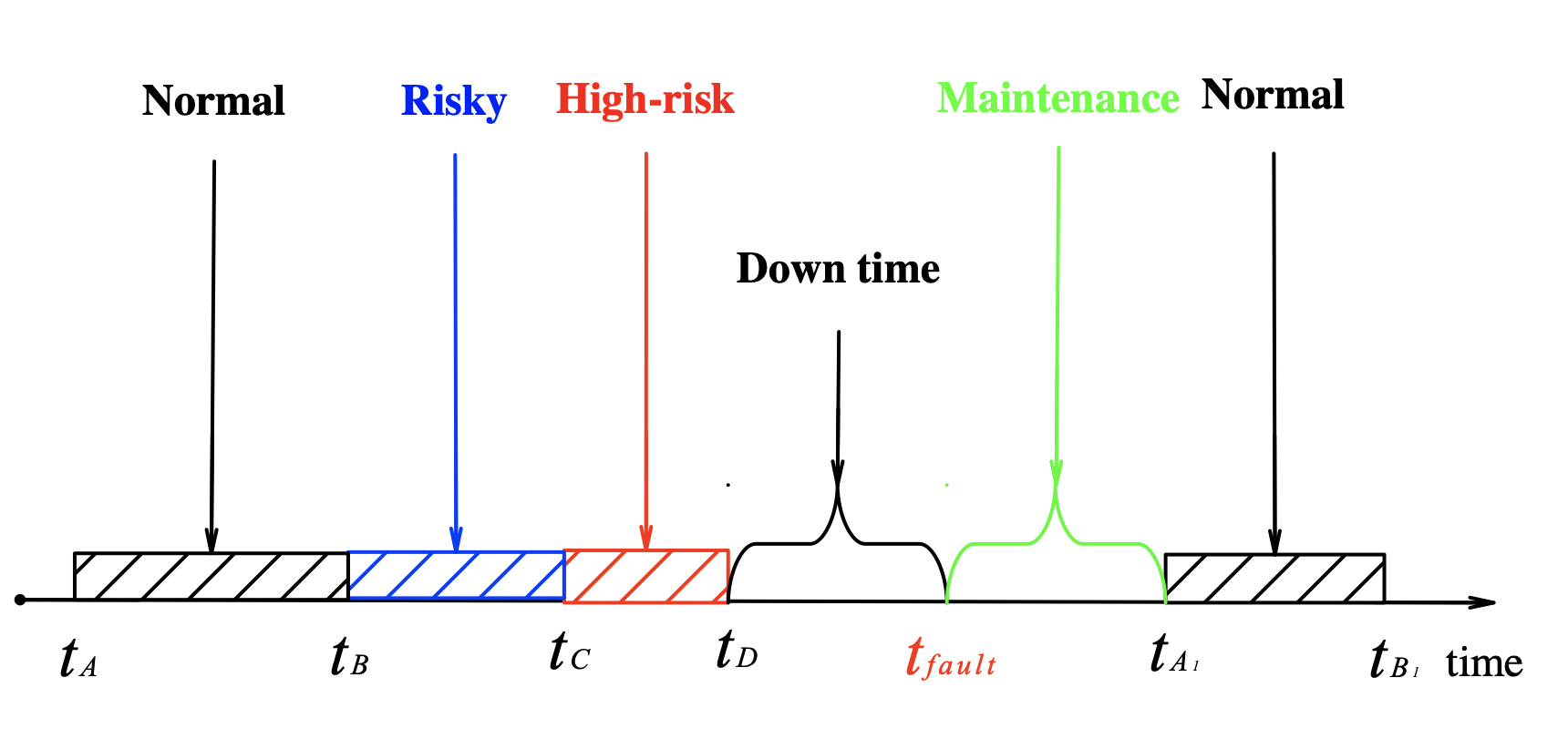}
\caption{The different intervals from normal to maintenance}
\label{fig:different_intervals}
\end{figure}

As Figure~\ref{fig:different_intervals} shows, we define the process of deterioration as risky and high-risk, then, a machine will usually experience three stages before the first down time: normal, risky, and high-risk. 

Assuming a machine starts running at time $t_A$; at time $t_B$, it turns to be risky; then, it enters to high-risk operation at time $t_C$, we define $t_D$ as the end of the high-risk interval, once we fail to detect or predict a fault at time $t_D$, the machine will experience the first breakdown from time $t_D$ to time $t_{fault}$, $t_{fault}$ is the end of down time. Then, $t_A<t_B<t_C<t_D<t_{fault}$. After maintenance, it will enter the next loop. The time period from normal operation to the beginning of the first breakdown including: the normal operation time interval $[t_A,t_B)$, the risky operation time interval $[t_B,t_C)$ and the high-risk operation time interval $[t_C,t_D]$.

Suppose we got $N$ observed signals of a machine in a normal-to-high-risk loop. Which including $N_1$ signals in the normal operation interval, $N_2$ signals in the risky operation interval and $N_3$ signals in the high-risk operation interval, therefore, $N =N_1+N_2+N_3$. 

Let $t_i$ represent the acquisition time of the machine's $i$th observed signal $\mathbf{x}_i$, $t_A\le t_i\le t_D$, $1\le i\le N$. Then, let $x_{ij}$ presents the $j$th feature of signal $\mathbf{x}_i$, where $i = 1,\cdots,N$ , and, $j = 1,\cdots,k$, $k$ is the total feature dimension. Then,  the $i$th observation $\mathbf{x}_\mathbf{i}= (x_{i1},\cdots,x_{ik})$.  Let’s define $y_i$ as the status of the machine, where $y_i \in{0, 1, 2}$, when $y_i=0$, the machine was normal, $y_i=1$ means it was risky and $y_i=2$ means high-risk status. The status set of the machine is $\mathbf{y} = (y_1,\cdots,y_N)$. Thus, a set of the signal features and states is defined as $X={(\mathbf{x}_1,y_1),\cdots,(\mathbf{x}_N,y_N)}$, which is also called labeled training data sets.

The fault prediction model $f(\cdot)$ can be built by machine learning algorithms, the predicted result of time $t_i$ is $\widehat{y_i}=f(\mathbf{x}_i)$. The prediction set of the whole interval is $\hat{\mathbf{y}}=\left(\widehat{y_1},\cdots,\widehat{y_N}\right)$, the target of the training is to make the errors between $\mathbf{y}$ and  $\hat{\mathbf{y}}$ as small as possible.

%% file: sections/cascade_OPAI.tex
\section{Cascade classification and Online Prediction Ability Index }\label{Sec:3}

\subsection{OPAI definition and demonstration}

The existing classification model evaluation metrics, such as accuracy, error rate, RSME, etc., are static metrics, which are suitable for evaluating offline models. Predicting the rotor parts fly-off faults requires continuous online monitoring, analyzing the time series collected from sensors, and classify the rotor's operating status and the degree of deterioration into labeled categories. The closer time to the failure, the more dangerous the rotor is. Therefore, the traditional offline static evaluation metrics are difficult to meet the demand. Thus, this paper proposes OPAI to evaluate the accuracy and consistency of a prediction model. The definition and calculation of OPAI will be described in the following paragraph.

\subsubsection{The time-dependent accuracy evaluation of OPAI}

According to the above definition, there are three different operating states of the machine before breakdown: normal operation, risky operation and high-risk operation. Therefore, the prediction accuracy will be constructed by the prediction accuracies of the above three intervals together, which is calculated as Equations~\eqref{eqn::S_1}and~\eqref{eqn::S_2}. When the machine is normal, the working condition has no effect on the deterioration, so the prediction accuracy weight of each sample in this time interval is the same. Once the machine starts to deteriorate, the closer to the downtime, the higher the degree of deterioration is, and the more dangerous the machine is, so the prediction accuracy weight of the sample in the risky and the high-risk operation interval should increase with time. Moreover, after entering the high-risk operation, the deterioration speed increased, so the prediction accuracy weight of the sample in the high-risk interval should increase faster than those samples in the risk operation interval.

\begin{strip}
\begin{gather}
    S = 1 - \sum_{i=1}^{N}{Error(i)}
    \label{eqn::S_1} \\ 
    Error(i)=\left\{\begin{matrix}
    \alpha\ ln(N_1) &  t_A\le t_i<t_B,\widehat{y_i}\neq y_i \\ 
    \beta\ ln(i-N_1)+\alpha\ ln(N_1) &  t_B\le t_i<t_C,\widehat{y_i}\neq y_i \\ 
    \gamma{(i-N_1-N_2)}^2+\beta ln(N_2)+\alpha\ ln(N_1) & t_C\le t_i\le t_D,\widehat{y_i}\neq y_i \\
    0 &  others 
    \end{matrix}\right.
   \label{eqn::S_2}
\end{gather}
\end{strip}

$\alpha$, $\beta$, $\gamma$ are constant adjustment coefficients, making $S\in[0, 1]$. We adjust the three parameters by making the normal, risky, and high-risk operation intervals account for 20\%, 30\% and 50\% of the total score.

\subsubsection{The consistence evaluation of OPAI }

The consistence evaluation index is calculated as Equations~\eqref{eqn::C_1}, ~\eqref{eqn::C_2},~\eqref{eqn::C_3} and ~\eqref{eqn::C_4}, the overall prediction stability $C$ is calculated by averaging the prediction stability of three intervals, where $C_1, C_2, C_3$ indicate the model prediction stability metrics corresponding to the normal interval, risky operation interval, and the high-risk operation interval. As the prediction model should classify the status of the rotor in a consecutive time series. The more consistent the predictions are, the higher the robustness of the model is, which has higher practical application values.

\begin{gather}
    C\ =\frac{1}{3}(C_1+C_2+C_3)
    \label{eqn::C_1} \\ 
    C_1\ =\ 1-\frac{1}{N_1-1}\sum_{t_i=t_A}^{t_{B-1}}{(\widehat{y_{i+1}}-\widehat{y_i})}^2
    \label{eqn::C_2}\\
    C_2\ =\ 1-\frac{1}{N_2-1}\sum_{t_i=t_{B+1}}^{t_{C-1}}{(\widehat{y_{i+1}}-\widehat{y_i})}^2
    \label{eqn::C_3}\\
    C_3\ =\ 1-\frac{1}{N_3-1}\sum_{t_i=t_{C+1}}^{t_{D-1}}{(\widehat{y_{i+1}}-\widehat{y_i})}^2
    \label{eqn::C_4}   
\end{gather}

\subsubsection{Index demonstration }

To verify the validity and rationality of OPAI, this paper will design two different scenarios, i.e., one failure and two failures, then calculate the accuracy, $S$ and $C$ scores of different prediction models in different scenarios. Accuracy is calculated as $accuracy = \frac{Correctly\ predicted\ number}{Total\ observation\ number}$ . We will discuss whether the $S$ and $C$ scores are valid and reasonable in such scenarios.

\begin{table}[h]
\caption{Scores of different failure scenarios}
\label{table:Scores_scenarios}
\centering
\begin{tabular}{p{3mm}p{6mm}ccccc}
\toprule
\textbf{Fault times} &          & \textbf{Model 1} & \textbf{Model 2} & \textbf{Model 3} & \textbf{Model 4} & \textbf{Model 5} \\ \midrule
\multirow{3}{*}{1}   & Accuracy & 0.8              & 0.8              & 0.8              & 0.8              & 0.8              \\
                     & $S$        & 0.6447           & 0.8303           & 0.8800           & 0.8572           & 0.7950           \\
                     & $C$        & 0.8632           & 0.5812           & 0.8718           & 0.6581           & 0.5983           \\  \midrule
\multirow{3}{*}{2}   & Accuracy & 0.8              & 0.8              & 0.8              & 0.8              & 0.8              \\
                     & $S$        & 0.6479           & 0.8052           & 0.8800           & 0.8550           & 0.7962           \\
                     & $C$        & 0.8421           & 0.4211           & 0.8158           & 0.5526           & 0.6140        \\
		\bottomrule
\end{tabular}
\end{table}

Table~\ref{table:Scores_scenarios} summarizes the scores of 5 prediction models in different scenarios, and the time series plots of prediction results of model 1-5 under two scenarios are shown in Figure~\ref{fig:MCMC_all} below. To better reflect whether $S$ and $C$ are valid and reasonable, we assume 5 models have same overall accuracies i.e. 80\%, but making the prediction accuracy of the five models vary widely in different operation intervals. Therefore, the scores of $S$ and $C$ are different due to the different weights in different operation intervals and the prediction stability of the models. The scores of different models will be discussed in conjunction with the time series plots of the prediction results.

\begin{figure*}
  \centering
  \subcaptionbox{The prediction results of model 1  \label{figure:MCMC_1}}
     {\includegraphics[width = .49\linewidth]{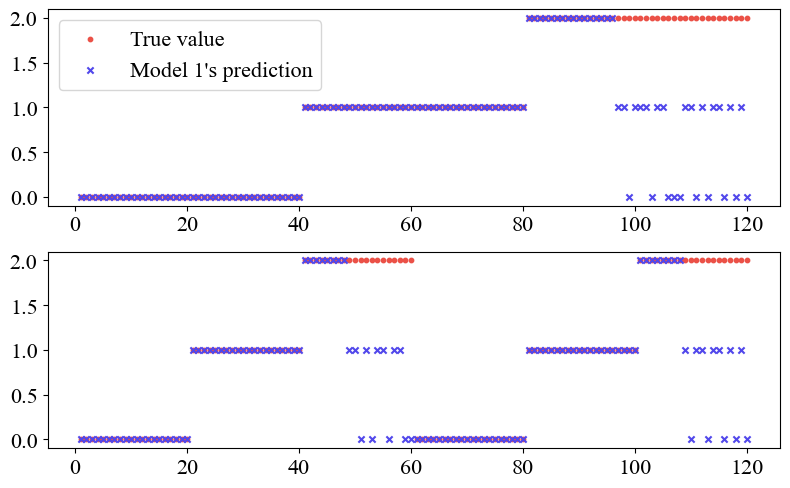}} 
  \subcaptionbox{The prediction results of model 2  \label{figure:MCMC_2}}
     {\includegraphics[width = .49\linewidth]{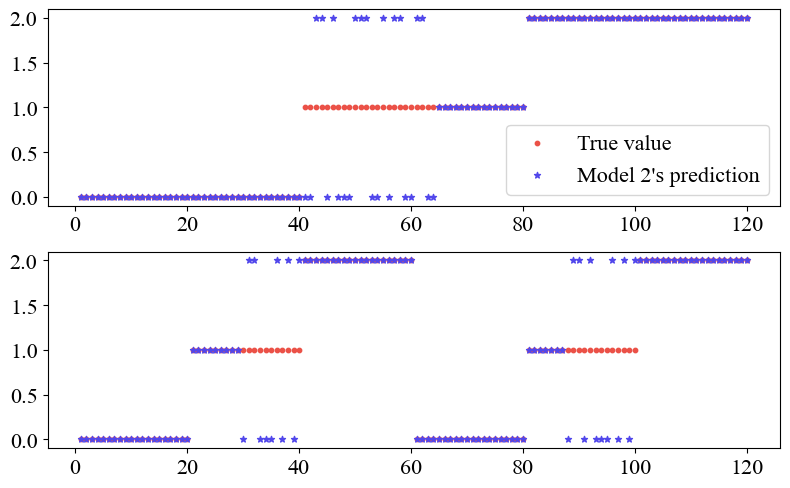}} \\
  \subcaptionbox{The prediction results of model 3  \label{figure:MCMC_3}}
     {\includegraphics[width = .49\linewidth]{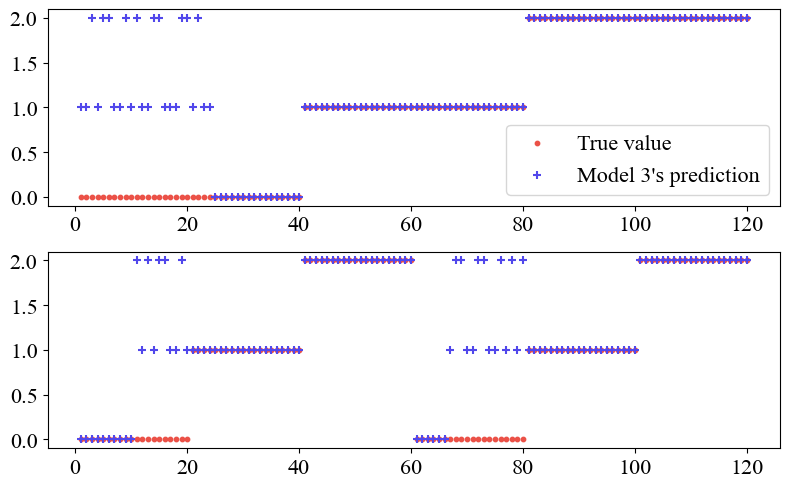}} 
  \subcaptionbox{The prediction results of model 4 \label{figure:MCMC_4}}
     {\includegraphics[width = .49\linewidth]{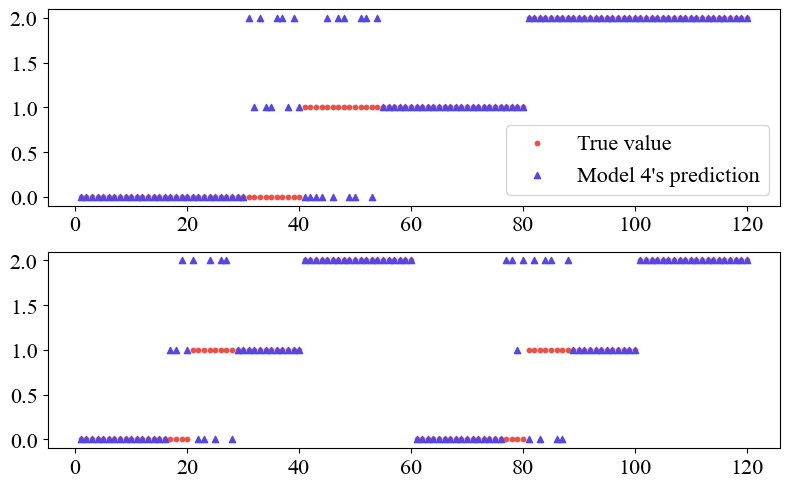}} \\
  \subcaptionbox{The prediction results of model 5 \label{figure:MCMC_5}}
     {\includegraphics[width = 0.7\linewidth]{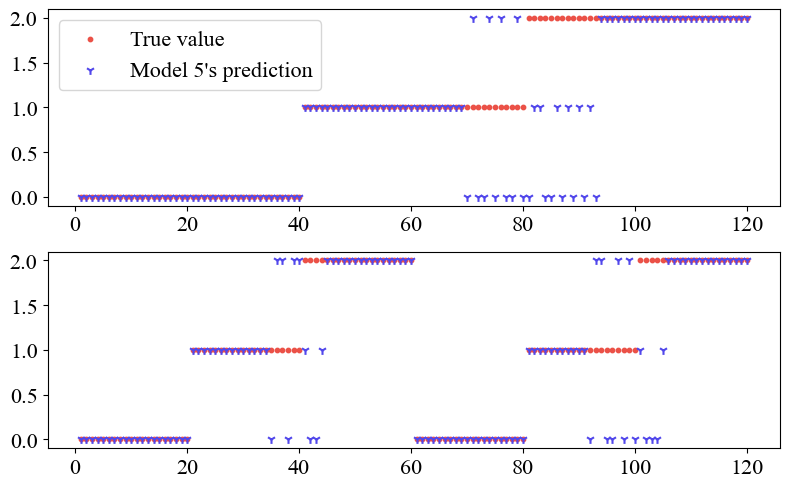}}\\
  \caption{The scatter plot of 5 models’ prediction results in the case of one and two failures scenarios}
  \label{fig:MCMC_all}
 \end{figure*}

Illustrated by Figure ~\ref{figure:MCMC_1} and Table~\ref{table:Scores_scenarios}, model 1 can predict accurately in normal and risk operation intervals, but has difficulty in identifying samples in high-risk intervals and missed high-risk samples in all scenarios. Therefore, even with 80\% prediction accuracy, model 1 has $S$ score of 0.6447 for one failure and $S$ score of 0.6479 for two failures, both of which are lower than accuracy, indicating that the metric $S$ pays more attention to the prediction accuracy of high-risk intervals. The stability evaluation index $C$ of model 1 is higher than 0.8 in both scenarios, indicating that model 1 has good stability once deployed in real-world application.

Looking through Figure ~\ref{figure:MCMC_2}, we find that model 2 can predict well in the normal and high-risk operation intervals, but performs poorly in the risky operation interval, which has slightly higher $S$ score of 0.8303 in the case of one failure and $S$ score of 0.8052 in the case of two failures. The risky operation interval is between the normal operation and high-risk operation intervals, which tends to be much more difficult to predict, and the $S$ score of the model 2 indicates its difficulty in identifying early deterioration trends. The erroneous predictions of model 2 fluctuate significantly between normal and high-risk, so the model’s stability score $C$ is low, 0.5812 in the case of one failure and 0.4211 in the case of two failures, which makes model 2 difficult to apply in reality.

Observing Figure ~\ref{figure:MCMC_3}, it is known that model 3 can accurately predict the fault deterioration trend, and  has false alarm in the normal operation interval. The false alarms fluctuate between risk and high-risk intervals. Since each sample in the normal operation interval has the same weight, the model has $S$ score of 0.88 for both scenarios, stability score $C$ is 0.8718 for one fault scenario, and 0.8158 for two-fault scenario. With higher $C$ and $S$ scores for both scenarios, the consistency is acceptable. The model is able to identify all samples in the risky and high-risk operation intervals, which is valuable for realistic guidance. However, its prediction in normal operation intervals is poor and too sensitive with many false alarms. Such false alarms may not cause damage to the equipment itself, but could lead to some misguidance for O \& M.

Observing Figure ~\ref{figure:MCMC_4}, we can find that model 4 can accurately predict in the high-risk interval, but cannot correctly identify all samples in the normal and risky intervals, the prediction results fluctuate greatly. The score of $S$ is 0.8572 for one fault scenario and 0.8550 for two-fault scenario. Although the score of $S$ is higher than the accuracy, score of $C$ is lower than 0.7, i.e., the score of $C$ is 0.6581 for one failure scenario and 0.5526 for two-failure scenario. Since we define normal, risky and high-risk by the time to the failure, the edge of normal interval may already show early deterioration characteristics. 

Observing Figure ~\ref{figure:MCMC_5}, it is known that model 5 has $S$ score of 0.7950 and $C$ score of 0.5983 for one failure scenario, and $S$ score of 0.7962 and $C$ score of 0.6140 for two-failure scenario. Model 5 can predict well in the normal operating interval, but perform poorly in the risky and high-risk operation intervals. It reflects the most difficult situation in reality, in the edge of the risky interval and the early part of the high-risk interval, the observed data may show complex patterns and unstable time series. If successfully identifying early warning from risky to high-risk transition, great economical benefits could be achieved. 

In summary, considering the prediction accuracy, the above 5 models have the same performance. However, the scores of $S$ and $C$ can evaluate the prediction model from different angles, the score of $S$ reflects the online prediction ability of the model and the score of $C$ reflects the prediction stability of the model. In practice, if the model fails to alarm, it will lead to greater loss. So the samples in the deterioration intervals should have high weights. Models with high $S$ score should be selected. As the $S$ scores of models 2, 3 and 4 are higher than 0.8, they are considered as the candidates. Furthermore, considering the score of $C$, model 3 performs the best in terms of stability.

\subsection{The cascade classification algorithm }

\begin{figure}[h]
\centering
\includegraphics[width=\linewidth]{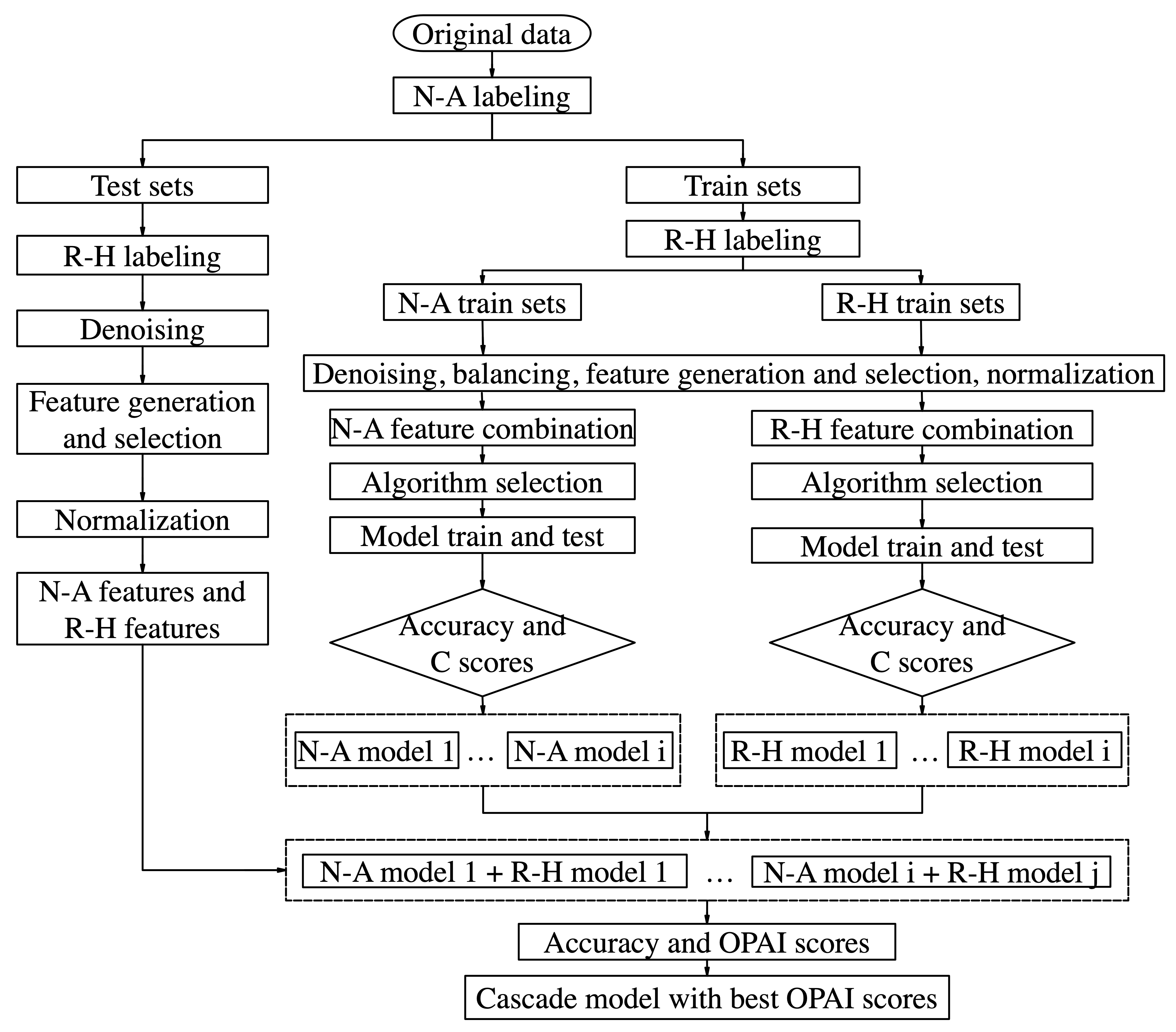}
\caption{The flow chart of cascade classifying}
\label{fig:flow_chart_cascade}
\end{figure}

We propose a cascade classification algorithm to predict the rotor parts fly-off deterioration process. The algorithm’s flow chart is shown in Figure ~\ref{fig:flow_chart_cascade}. After the original data is obtained, N-A labeling is performed. Samples in the normal intervals $[t_A,t_B)$ are labeled as Normal, samples in deterioration intervals $[t_B,t_D]$ are labeled as Abnormal. Then stratified sampling was used to select 80\% of the samples as the train sets and 20\% of the samples as the test sets. For the fault machine, R-H labeling is performed. Samples in risky interval $[t_B,t_C)$ are labeled as Risky, samples in high-risk intervals are labeled as High-risk. For the rotor parts fly-off faults, $t_B$ is 1 month from the time of failure, $t_C$ is 1day from the time of failure. Then, 80\% of the N-A samples are used to train the N-A classifier, and 80\% of the R-H samples are used to train the R-H classification model. Then, wavelet de-noising and under-sampling are applied to the train sets, then we extract the time domain and frequency domain features of signals. The effective feature combination is selected and standardized. At the same time, de-noising, feature engineering and standardization are applied to test sets.

After determining the feature combination, it is necessary to select an appropriate machine learning algorithm. The N-A and R-H train sets are used to train to the binary classifiers respectively. The binary accuracy and $C$ score of the five-fold cross validation for each algorithm are calculated. Then, models with high binary accuracy and $C$ score are selected as candidates. Model combinations of cascade classification are constructed. Finally, the combined models are tested with the test sets. The accuracy and OPAI scores of the test sets are calculated. The model combination with the highest scores is selected as the best option. 

%% file: sections/experiment.tex
\section{Experiments}\label{Sec:4}

In this section, vibration signal data from rotors of compressors and turbines are collected to verify the cascade classifying algorithm and OPAI. 

\subsection{Data description}

\begin{figure}[h]
\centering
\includegraphics[width=\linewidth]{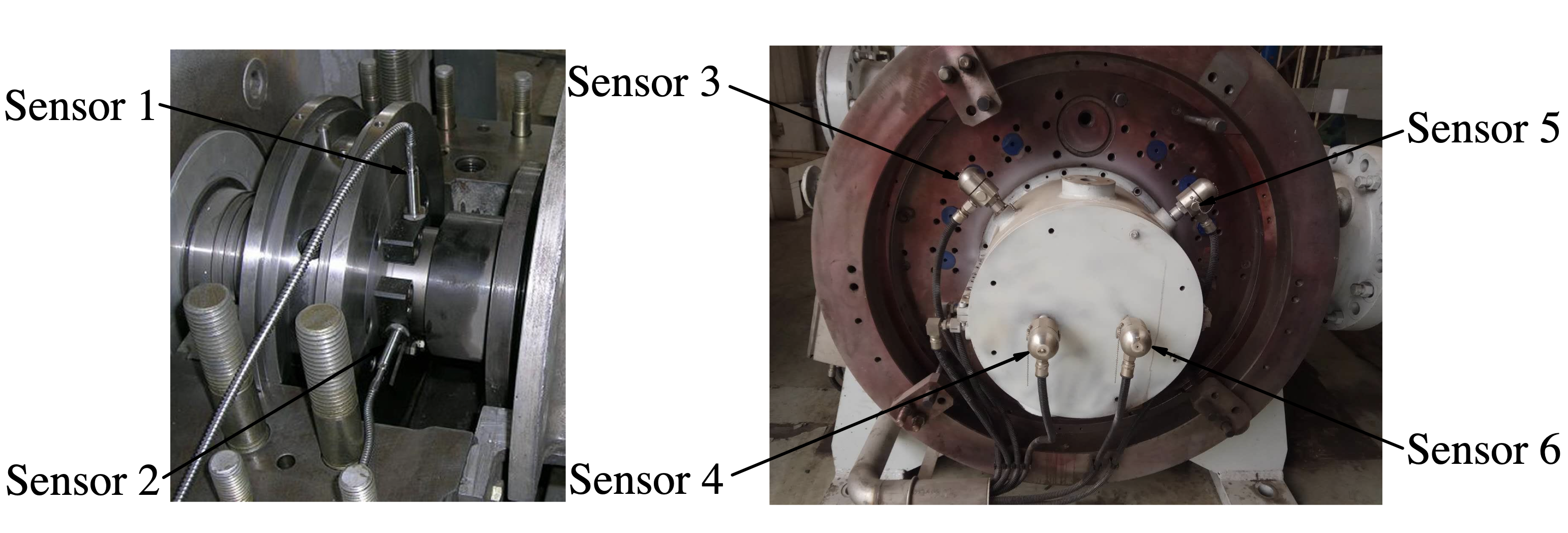}
\caption{Diagram of rotor vibration sensor distribution}
\label{fig:sensors}
\end{figure}

As shown in Figure ~\ref{fig:sensors}, two eddy current displacement sensors (Sensor1 and Sensor 2) are arranged in the bearings of the rotor at both ends of the large turbine rotating machine to measure the radial vibration of the rotor. Another 4 eddy current displacement sensors (Sensor3, Sensor4, Sensor5 and Sensor6) are arranged at the shaft end to measure the axial vibration. The signals are connected to SG8000 data collector for data adjustment, acquisition and storage. In this paper, the vibration signal data of 14 rotating machines were collected for 6 months, among which, 7 machines had one failure and 7 machines had been operating normally for 6 months. The data of 6 sensors of each machine were collected basically at the same time. Due to the limitation of the data source, instead of continuous monitoring and collecting vibration data for 6 months, data is extracted at certain time intervals. For example, for machines with one failure, the training and testing samples are extracted from 6 months, 3 months, 1 month, 1 week, 1 day, 1 hour and minutes before the failure. For normal machine, the training and testing samples are extracted from 6 months, 3 months, 1 month, 1 week, 1 day, 1 hour, and minutes before the last observation time. 

Tables ~\ref{table: sample_size} and~\ref{table: sample_test}  summarize the vibration wave collected by the 6 sensors for the 14 machines. Machine 1, 2, 3, 4, 5, 6 and 7 have one failure. Machine 8, 9, 10, 11, 12, 13, 14 are normal. Therefore, Machine 1 to Machine 7 have signals in the risky and high-risk intervals, but all signals  for Machine 8 to Machine 14 are in normal interval. In the Tables ~\ref{table: sample_size} and ~\ref{table: sample_test}, 6 vibration waves collected by the 6 sensors form a wave array. It is obvious that most waves are collected during the normal intervals, from 6 months to 1 month before breakdown of the faulty machine and all operational time periods of the normal machine. There are 13,203 normal wave arrays in the train sets and 5,368 normal wave arrays in the test sets, which account for about 70\% of the total wave arrays. The number of normal wave arrays is much higher than the number of wave arrays in the other intervals. Furthermore, for most of the faulty machines, the number of wave arrays in the risky interval (1 week to the 1 day before the failure) is slightly higher than the number of wave arrays in the high-risk interval (1 hour and minutes before the failure). Therefore, before training the classifiers, we have to equalize the training samples in different intervals.

\begin{table*}[t]
\caption{Samples size of the vibration waves for each machine in the train sets at different intervals}
\label{table: sample_size}
\centering
\setlength{\tabcolsep}{12pt}
\begin{tabular}{@{}cccccccccccc@{}}
\toprule
\textbf{Label}     & \textbf{M1} & \textbf{M2} & \textbf{M3} & \textbf{M4} & \textbf{M5} & \textbf{M8} & \textbf{M9} & \textbf{M10} & \textbf{M11} & \textbf{M12} & \textbf{All} \\ \midrule
\textbf{High-risk} & 798         & 599         & 501         & 472         & 500         & 0           & 0           & 0            & 0            & 0            & 2870         \\
\textbf{Risky}     & 623         & 609         & 574         & 611         & 574         & 0           & 0           & 0            & 0            & 0            & 2991         \\
\textbf{Normal}    & 1021        & 766         & 860         & 996         & 866         & 1744        & 1740        & 1737         & 1737         & 1736         & 13203        \\   \midrule
\textbf{All}       & 2442        & 1974        & 1935        & 2079        & 1940        & 1744        & 1740        & 1737         & 1737         & 1736         & 19064        \\ \bottomrule
\end{tabular}
\end{table*}

\begin{table}[h]
\caption{Samples size of the vibration waves for each machine in the test sets at different intervals}
\label{table: sample_test}
\centering
\setlength{\tabcolsep}{12pt}
\begin{tabular}{@{}cccccc@{}}
\toprule
                   & \textbf{M6} & \textbf{M7} & \textbf{M13} & \textbf{M15} & \textbf{All} \\ \midrule
\textbf{High-risk} & 305         & 311         & 0            & 0            & 616          \\
\textbf{Risky}     & 649         & 587         & 0            & 0            & 1236         \\
\textbf{Normal}    & 957         & 941         & 1743         & 1727         & 5368         \\  \midrule
\textbf{All}       & 1911        & 1839        & 1743         & 1727         & 7220         \\ \bottomrule
\end{tabular}
\end{table}

Table ~\ref{table: sample_description} shows the maximum, minimum, mean and standard deviation of the vibration waves in different intervals. We can observe that the standard deviation of the vibration waves in the normal interval is much smaller than others, for example, the standard deviation of waves in normal interval for sensor 1 is 2.0366, which is smaller than 6.7985 of waves in risky interval and 11.1055 of waves in high-risk interval. Possible reason is that the rotor vibrates smoothly in the normal interval. 

\begin{table}[h]
\caption{Description of vibration waves collected by 6 sensors in different intervals}
\label{table: sample_description}
\centering
\setlength{\tabcolsep}{12pt}
\begin{tabular}{@{}ccccc@{}}
\toprule
                                   &               & \textbf{Normal} & \textbf{Risky} & \textbf{High-risk} \\ \midrule
\multirow{4}{*}{\textbf{Sensor 1}} & \textbf{Max}  & 16.1398         & 28.6255        & 40.4576            \\
                                   & \textbf{Min}  & -12.1615        & -26.3996       & -32.3692           \\
                                   & \textbf{Mean} & 0.0005          & 0.0057         & 0.0000             \\
                                   & \textbf{Std.} & 2.0366          & 6.7985         & 11.1055            \\  \midrule
\multirow{4}{*}{\textbf{Sensor 2}} & \textbf{Max}  & 19.9629         & 24.1449        & 44.7034            \\
                                   & \textbf{Min}  & -29.1171        & -23.1815       & -43.0939           \\
                                   & \textbf{Mean} & 0.0000          & 0.0057         & 0.0000             \\
                                   & \textbf{Std.} & 2.8504          & 6.8678         & 12.3792            \\  \midrule
\multirow{4}{*}{\textbf{Sensor 3}} & \textbf{Max}  & 10.5867         & 41.0726        & 42.1692            \\
                                   & \textbf{Min}  & -8.5252         & -40.7834       & -41.9020           \\
                                   & \textbf{Mean} & 0.0000          & 0.0000         & 0.0000             \\
                                   & \textbf{Std.} & 2.1308          & 10.1692        & 12.5702            \\  \midrule
\multirow{4}{*}{\textbf{Sensor 4}} & \textbf{Max}  & 12.8826         & 39.0154        & 36.7583            \\
                                   & \textbf{Min}  & -10.7369        & -39.4274       & -38.3679           \\
                                   & \textbf{Mean} & 0.0000          & 0.0000         & 0.0000             \\
                                   & \textbf{Std.} & 2.2505          & 9.9623         & 11.8240            \\  \midrule
\multirow{4}{*}{\textbf{Sensor 5}} & \textbf{Max}  & 19.0014         & 45.9220        & 54.2757            \\
                                   & \textbf{Min}  & -28.0415        & -38.3638       & -45.8510           \\
                                   & \textbf{Mean} & 0.0000          & 0.0000         & 0.0000             \\
                                   & \textbf{Std.} & 4.8239          & 12.6074        & 15.6947            \\   \midrule
\multirow{4}{*}{\textbf{Sensor 6}} & \textbf{Max}  & 22.7479         & 43.7437        & 55.1661            \\
                                   & \textbf{Min}  & -31.7740        & -36.0787       & -49.0554           \\
                                   & \textbf{Mean} & 0.0000          & 0.0000         & 0.0000             \\
                                   & \textbf{Std.} & 5.0908          & 12.1179        & 16.1177            \\ \bottomrule
\end{tabular}
\end{table}

The absolute maximum and minimum of vibration waves in the normal interval are also smaller than the risky and high-risk intervals, e.g., for sensor 1, the maximum of normal waves is 16.1398, which is smaller than 28.6255 of risky waves and 40.4576 of high-risk waves. It indicates the amplitude of the vibration waves in normal interval is smaller than others. The reason could be that the radial and axial displacements of the rotor fluctuate within a reasonable range. As deterioration begins, the amplitude and standard deviation of vibration waves increase. The amplitudes of sensors 1, 2, 5, and 6 are slightly lower in the risky interval (sensor 1 $[-26.3996, 28.6255]$, sensor 2 $[-23.1815, 24.1449]$, sensor 5 $[-38.3638, 45.9220]$, sensor 6 $[-36.0787, 43.7437]$) than in the high-risk interval (sensor 1 $[-32.3692, 40.4576]$, sensor 2 $[-43.0939, 44.7034]$, sensor 5 $[-45.8510, 54.2757]$, sensor 6 $[-49.0554, 55.1661]$). 

There is no much difference for the amplitudes of sensors 3 and 4 in the risky and high-risk intervals, e.g., for sensor 3, the maximum value of waves for risky and high-risk are 41.0726 and 42.1692, the minimum value of waves for risky and high-risk are -40.7834 and -41.9020. One reason could be that the radial vibration will increase slowly as the deterioration begins, but the axial displacement may already fluctuate at early stage, which poses a great challenge to the R-H classification.

\begin{figure}
\centering
\includegraphics[width=\linewidth]{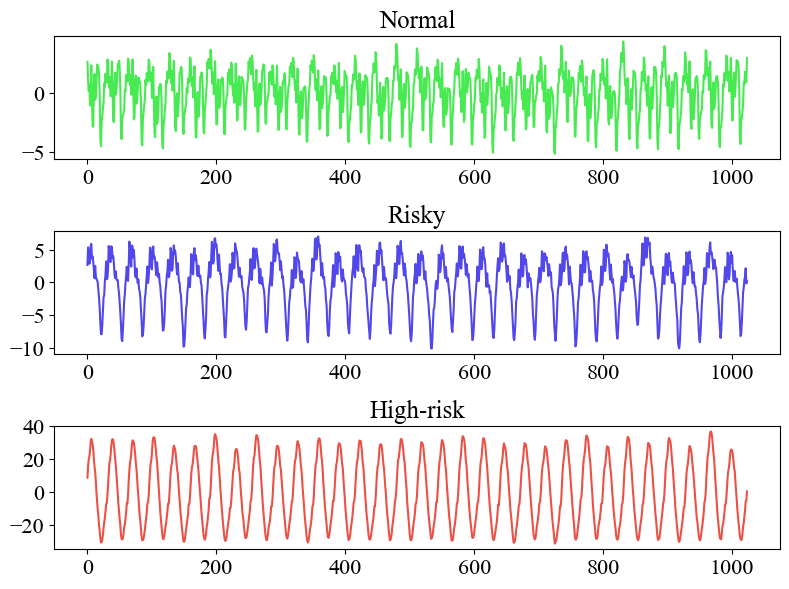}
\caption{Samples of vibration waves in different intervals}
\label{fig:vibration_samples}
\end{figure}

We select vibration waves from sensor 1 of Machine 1 in three intervals for visualization. The time series of vibration wave samples are shown in Figure ~\ref{fig:vibration_samples}. A wave includes 1024 samples. There is no much difference for the amplitude and shape of waves between the risky (blue) and normal intervals (green), but the vibration frequency changes obviously. As deterioration increases, the vibration frequency decreases obviously (from normal-green to risky-blue, high-risk-red). Thus, features in frequency domain can be extracted as inputs to the classifiers.

\subsection{The experiment process}

\begin{figure}
\centering
\includegraphics[width=\linewidth]{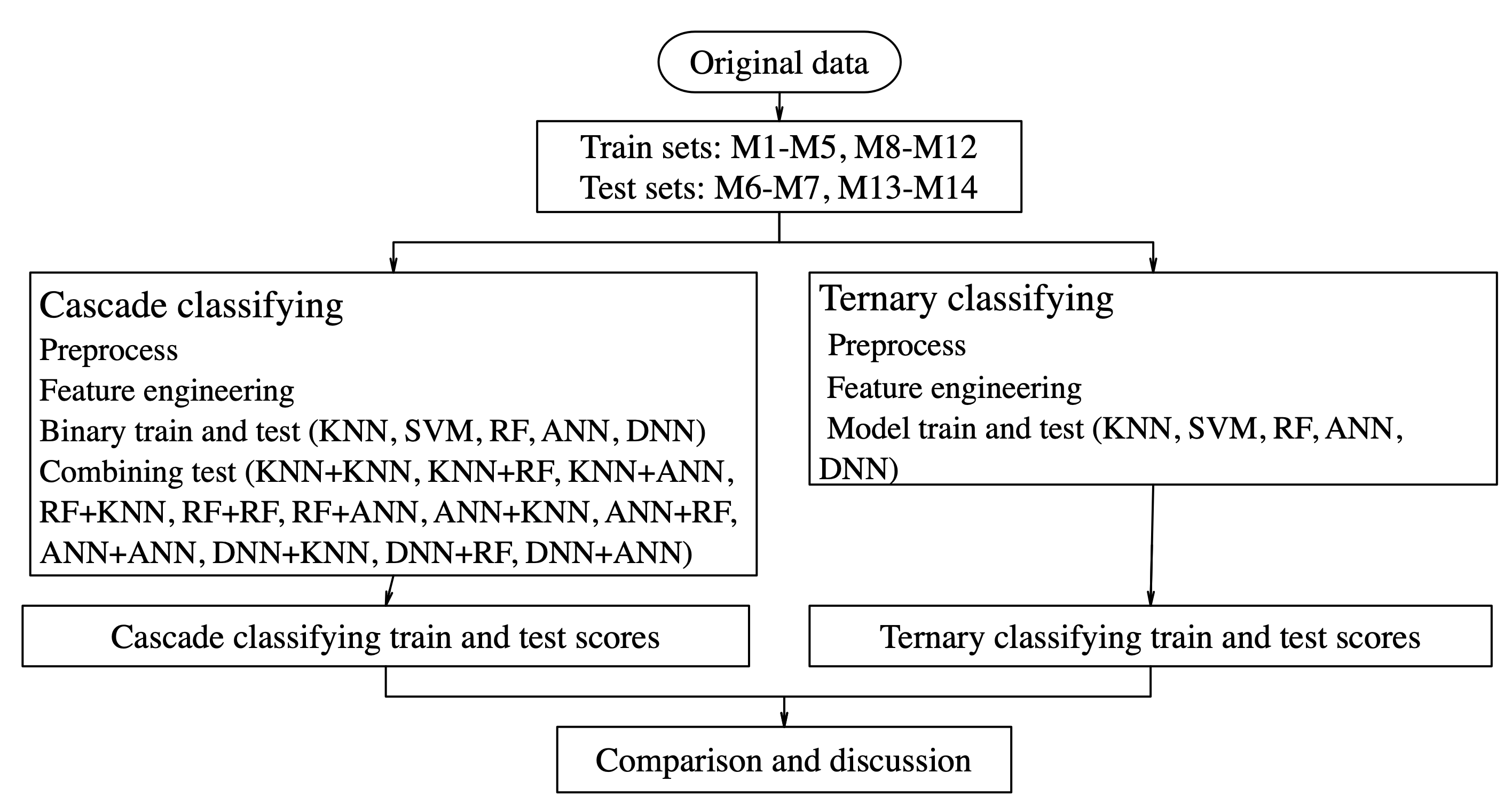}
\caption{The flow chart of computational experiment}
\label{fig:flow_chart}
\end{figure}

To verify the effectiveness of the cascade classifying process and OPAI, a comparative analysis is conducted. The process of the computational experiment is shown in Figure ~\ref{fig:flow_chart}. The original data is collected from 14 machines. First, we split the original data into train sets and test sets. Data in the train sets are collected from 5 faulty machines with one failure (M1-5) and 5 normal machines (M8-12). Data in the test sets are collected from 2 faulty machines with one failure (M6-7) and 2 normal machines (M13-14). Then, we perform preprocess for cascade and ternary classifying, including labeling, wavelet denoise, and sample equalization. Then, features in the time domain and time-frequency domain are extracted, appropriate feature combinations are selected for different classifiers. Five machine learning algorithms, KNN, SVM, RF, ANN and DNN, are applied for the cascade and ternary classification. 

For cascade classification, KNN, SVM, RF, ANN and DNN are applied for N-A and R-H classification. We use N-A training sets to train the N-A classifiers and R-H training sets to train the R-H classifiers, calculate the 5-fold cross validation scores respectively. Since the KNN, RF and ANN classifiers get higher training scores in this binary classification (as shown in Table ~\ref{table:scores_of_5fold_train}), we further select these 3 simple classifiers to construct 9 N-A + R-H cascade classifiers (as shown in the middle left of the Figure ~\ref{fig:flow_chart} flowchart). Then, test sets are used to validate these 9 cascade classifiers, the training and testing scores are calculated.

For ternary classification experiment, 5 machine learning algorithms KNN, SVM, RF, ANN and DNN are directly applied to train and test sets. The training and testing scores are calculated respectively and shown in Table ~\ref{table:scores_of_ternary_classifiers}.

\subsubsection{Feature extraction}

Time domain statistical features, such as, root mean square (RMS), mean, variance, skewness, kurtosis and crest factor are extracted from the cleaned signals. We extracted 12 features of the denoised waves $x(m)$, the features and calculation formulas are listed in Table ~\ref{table:statistical_features_extracted}.

\begin{table}[h]
%\tiny
		\caption{Time domain features extracted from denoised waves}
		\label{table:statistical_features_extracted}
		\centering
		\begin{tabular} {cc} \toprule
		\textbf{Features} 		&	\textbf{Formula}   \\  \midrule
		Maximum 		&	$X_{\text{max}}=\max \left( x(m) \right)$ 	\\ [5pt]  
		Minimum 		&	$X_{\text{min}}=\min \left( x(m) \right)$ 	\\ [5pt]  
		Mean 		&	$X_{\text{mean}}=\frac{1}{M} \sum_{m=1}^{M}{x(m)}$ 	\\ [5pt]
		Peak to peak 	&	$X_{\text{pp}} = X_{\text{max}}- X_{\text{min}}$ 	\\ [5pt]
		Variance 		&	$X_{\text{var}}=\frac{1}{M} \sum_{m=1}^{M}{\left(x(m) - X_{\text {mean }}\right) ^2}$ 	\\ [5pt]
		Standard deviation 		&	$X_{\text{std}}=\sqrt{\frac{1}{M} \sum_{m=1}^{M}\left( x(m) - X_{\text {mean }}\right) ^2}$ 	\\ [5pt]
		RMS 		&	$X_{\text{RMS}}=\sqrt{\frac{1}{M} \sum_{m=1}^{M}\left( x(m) \right) ^2}$ 	\\ [5pt]
		Skewness 		&	$X_{\text{skew}}= \frac{ \frac{1}{M} \sum_{m=1}^{M}\left( x(m) - X_{\text {mean }}\right) ^3}{X_{\text {var }}^{\frac{3}{2}}}$	\\ [5pt]
		Kurtosis 		&	$X_{\text{kurt}}= \frac{ \frac{1}{M} \sum_{m=1}^{M} \left( x(m) - X_{\text{mean}} \right) ^4}{ X_{\text{var}}^{2}}$ 	\\ [5pt]
		Shape factor 		&	$X_{\text{sf}}=\frac{X_{\text{RMS}}}{\left | X_{\text {mean }} \right | }$ 	\\ [5pt]
		Crest factor 		&	$X_{\text{cf}}=\frac{X_{\text{max}}}{X_{\text{RMS}}} $ 	\\ [5pt]
		Pulse factor 		&	$X_{\text{pf}}=\frac{X_{\text{max}}}{\left |  X_{\text {mean }} \right |  }$ 	\\ [5pt]
		Clearance factor 		&	$X_{\text{clear}}=\frac{X_{\text{max}}}{\left(\frac{1}{M} \sum_{m=1}^{M} \sqrt{\left| x(m) \right|}\right)^2}$ 	\\ [5pt]
		\bottomrule
	\end{tabular}
\end{table}

Time domain features are effective when the signal is smooth. But the rotating machinery vibration signals are nonlinear and unstable. To cope with these problems, Fast Fourier Transform (FFT) and WT are applied to extract the time-frequency domain features. We extract the amplitude characteristics of the FFT, the energy value and energy ratio of wavelet coefficients in each layer after decomposing the wave into 4 layers. The features are listed in Table~\ref{table:frequency_features_extracted}.

\begin{table*}[h]
\caption{Time-frequency domain features extracted from denoised waves}
\label{table:frequency_features_extracted}
\centering
\setlength{\tabcolsep}{30pt}
\begin{tabular} {cc} \toprule
		\textbf{Features}                        & \textbf{Description}                                                                     \\  \midrule
		Maximum amplitud                & Maximum amplitude based on FFT                                                  \\
		Mean amplitude                  & Minimum amplitude based on FFT                                                  \\
		Standard deviation of amplitude	 & 	Standard deviation of amplitude based on FFT                                    \\
		$\text{Energy}_{i} $                      & The energy value of the wavelet coefficient of the $i \text{th}$ layer                   \\
		$\text{Energy}_{i}   \text{ratio}$                 & The proportion of the energy value of the wavelet coefficient of the $i \text{th}$ layer \\
		\bottomrule
	\end{tabular}
\end{table*}

\subsubsection{Feature selection}

We extract 24 features from vibration waves of each sensor, among which, there are 13 time-domain features and 11 time-frequency domain features. In total, 144 features are extracted for each machine. Too many features cause dimension disaster. Proper feature selection is needed for different classifiers. 

Firstly, the features with coefficient of variation smaller than 1 should be deleted because they contain little information. 36 features are deleted, including kurtosis, crest factor, clearance factor, mean amplitude, energy ratio of the first and second layers for all sensors. Maximum and minimum, peak-to-peak for sensor 5 and sensor 6, maximum of sensor 2 are also deleted. In total, 43 features are excluded for further experiments.

Secondly, RF feature importance ranking is applied to evaluate the contribution of features. Features with Gini-index less than 0.01 are deleted. In total, we delete 74 features for N-A classification and 58 features for R-H classification respectively. 

Lastly, left features are visualized through boxplots. We prefer features with good inter-group difference and intra-group consistency. We identify 12 input features for each classifiers. The boxplots of features are shown in Figures ~\ref{fig:boxplot_N-A} and ~\ref{ fig:boxplot_R-H}. The feature name and explanation are listed in Appendix Tables ~\ref{table: N-A_inputs}and ~\ref{table: R-H_inputs}. In Figures  ~\ref{fig:boxplot_N-A} and ~\ref{fig:boxplot_R-H}, features distribute differently between Normal-Abnormal datasets and Risky-High-risk datasets. For example, the median and variance of $Std_{S1}$ (Figure  ~\ref{fig:boxplot_N-A}) in Normal group are obviously different from Abnormal group, which indicates it could classify the Normal and Abnormal condition.

\begin{figure}[ht]
    \centering
    \includegraphics[width=1\linewidth]{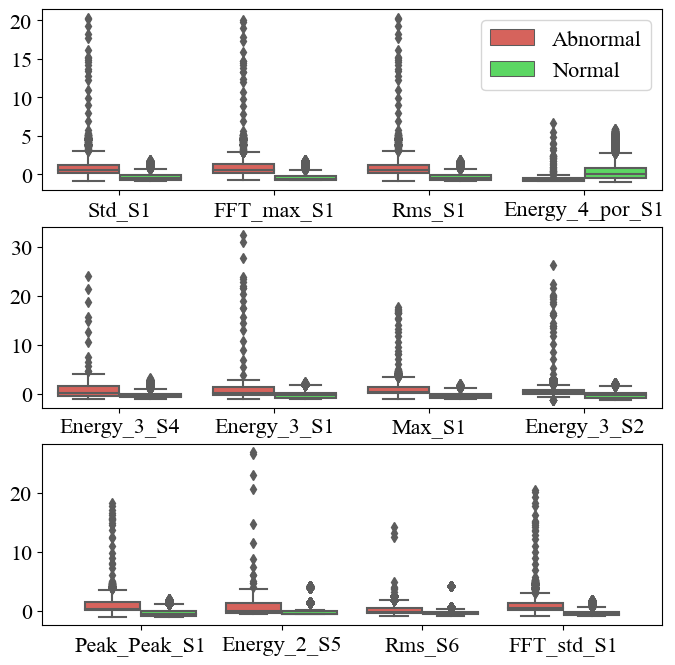}
    \caption{Boxplots of input features for N-A classifiers (left: Abnormal, right: Normal) }
    \label{fig:boxplot_N-A}
\end{figure}

\begin{figure}[ht]
    \centering
    \includegraphics[width=1\linewidth]{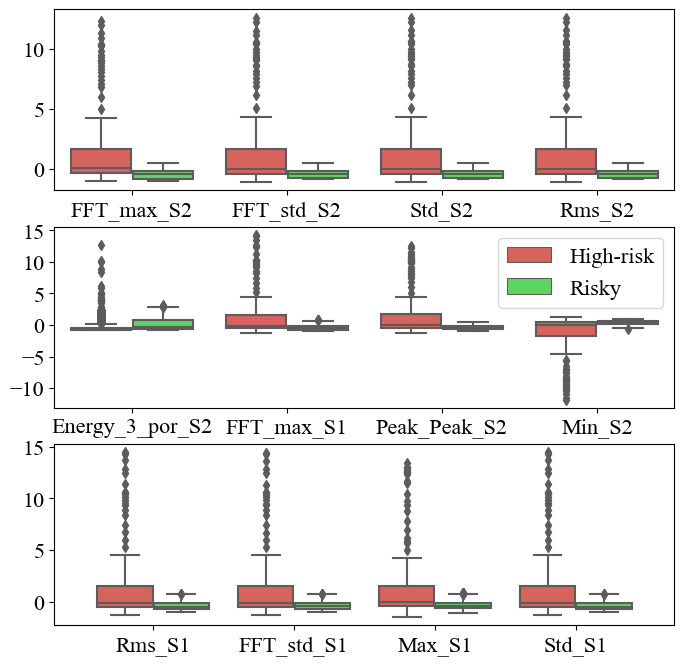}
    \caption{Boxplots of input features for R-H classifiers (left: High-risk, right: Risky)}
    \label{fig:boxplot_R-H}
\end{figure}

The feature selection process for ternary classification is the same as the cascade classification. First, 43 features with coefficient of variation less than 1 are deleted; Then, 67 features are deleted based on RF importance assessment; Finally, 14 features are selected as inputs for the ternary classifiers, shown in Appendix Figure ~\ref{fig:ternary_features}. The feature name and description are listed in the Appendix Table ~\ref{table: ternary_inputs}.

\subsection{Training and Testing }

KNN, SVM, RF, ANN and DNN are classical machine learning algorithms, which are widely used in the field of fault diagnosis and prediction. Among them, KNN is the simplest and can be used as benchmark model. The others are relatively complex and have their own advantages, e.g., DNN has the advantage of automated feature extraction and performs better in modeling and prediction for large-scale data. In this paper, what we are interested is not the algorithm comparison, but whether the proposed cascade classification process is effective. 

For ternary classification, KNN, SVM, RF, ANN and DNN are used to train and test. The 5-fold cross validation training and testing scores are calculated and shown in Table ~\ref{table:scores_of_ternary_classifiers}.

For cascade classification, N-A train sets are applied to train the N-A classifiers and R-H train sets are applied to train the R-H classifiers. We calculate the 5-fold cross validation scores respectively and show the results in Table ~\ref{table:scores_of_5fold_train}. The KNN, RF and ANN classifiers get higher training scores in each binary classification. These 3 simple classifiers are selected to construct 9 N-A + R-H cascade classifiers. 

Since N-A classifier and R-H classifier are trained separately, they do not consider the actual cascade classifying scenario, where N-A classifier’s outputs are fed into the subsequent R-H classifier. Therefore, we use total train sets to test the combined N-A+R-H classifiers, which gives us more convincing training scores and simulates the real-world classification. 9 model combinations are applied to the train and test sets. The scores are shown in Table 9. 

\subsection{The ternary classification model’s performance}

Table ~\ref{table:scores_of_ternary_classifiers} summarizes the training and testing scores of all ternary classifiers. In terms of consistency, the training $C$ scores of KNN and RF are 1 and 0.9772 respectively, which are higher than the $C$ scores of SVM (0.7683), ANN (0.7344), and DNN (0.7767). Testing $C$ scores of KNN and RF are 1 and 0.9737 respectively and showing similar performance as they are in train sets. So we can conclude that KNN and RF are more consistent in ternary classification. 

In terms of accuracy and $S$ score, KNN has the 0.944 training accuracy and 0.9462 S score. However, the test accuracy and $S$ score of KNN are 0.4537 and 0.4753, which is much lower than its training accuracy and $S$ score. It indicates that KNN has poor generalization ability. 

The test accuracy and $S$ score of RF are 0.6677 and 0.6750, which are the highest among the 5 models. The training accuracy and $S$ score of RF are 0.8677 and 0.8769, which is closed to the best models. As a result, we think RF is the best ternary classifier.

\begin{table*}[]
\caption{The training and testing scores of ternary classifiers}
\label{table:scores_of_ternary_classifiers}
\centering
\setlength{\tabcolsep}{20pt}
\begin{tabular}{@{}ccccccc@{}}
\toprule
\textbf{Dataset}                                  &          & \textbf{KNN}    & \textbf{SVM}    & \textbf{RF}     & \textbf{ANN}    & \textbf{DNN}    \\ \midrule
\multirow{3}{*}{T\textbf{rain (M1-M5, M9-M12)}}   & \textbf{Accuracy} & 0.9440 & 0.8256 & 0.8677 & 0.9136 & 0.8849 \\
                                         & \textbf{$S$}        & 0.9462 & 0.8405 & 0.8769 & 0.9103 & 0.8833 \\
                                         & \textbf{$C$}        & 1.0000 & 0.7683 & 0.9772 & 0.7344 & 0.7767 \\  \midrule
\multirow{3}{*}{\textbf{Test (M6, M7, M13, M14)}} & \textbf{Accuracy} & 0.4537 & 0.6217 & 0.6677 & 0.4582 & 0.6131 \\
                                         & \textbf{$S$}        & 0.4753 & 0.6196 & 0.6750 & 0.3978 & 0.6206 \\
                                         & \textbf{$C$}        & 1.0000 & 0.7889 & 0.9737 & 0.9993 & 0.9531 \\ \bottomrule
\end{tabular}
\end{table*}

\subsection{Proposed model’s performance}

The 5-fold cross validation scores of N-A classifiers and R-H classifiers are summarized in Table ~\ref{table:scores_of_5fold_train}. We can observe that the training accuracy and C score of all N-A classifiers are higher than 0.9, which indicate good prediction accuracy and consistency. The accuracy of KNN (0.9867), RF (0.9575) and ANN (0.9751) are higher than SVM (0.9136) and DNN (0.9456), they are selected as candidates for first stage binary classifiers. 

For R-H classifying, the training accuracy and $C$ scores of all R-H classifiers are slightly lower than N-A classifiers. The reason is that the deterioration process is unstable and nonlinear, it is hard to find a perfect classifier. 

$C$ scores of all 5 R-H classifiers are higher than 0.8, which meets the consistency requirement. The accuracy of KNN (0.8824), RF (0.8267) and ANN (0.8219) are higher than SVM (0.7167) and DNN (0.7427), they are chosen as candidates for the second stage classifiers.

\begin{table*}[]
\caption{The 5-fold cross validation training scores of the N-A and R-H classifiers}
\label{table:scores_of_5fold_train}
\centering
\setlength{\tabcolsep}{12pt}
\begin{tabular}{@{}cccccccc@{}}
\toprule
\textbf{Task}                             & \textbf{Data}                                            & \textbf{}         & \textbf{KNN} & \textbf{SVM} & \textbf{RF} & \textbf{ANN} & \textbf{DNN} \\ \midrule
\multirow{2}{*}{\textbf{N-A classifying}} & \multirow{2}{*}{\textbf{M1-M5, M8-M12}}                  & \textbf{Accuracy} & 0.9867       & 0.9136       & 0.9575      & 0.9751       & 0.9456       \\ \cmidrule(l){3-8} 
                                          &                                                          & \textbf{$C$}        & 0.9813       & 0.9926       & 0.9815      & 0.9741       & 0.9697       \\ \midrule
\multirow{2}{*}{\textbf{R-H classifying}} & \multirow{2}{*}{\textbf{Abnormal observations of M1-M5}} & \textbf{Accuracy} & 0.8824       & 0.7167       & 0.8267      & 0.8219       & 0.7427       \\ \cmidrule(l){3-8} 
                                          &                                                          & \textbf{$C$}        & 0.8284       & 0.9973       & 0.8598      & 0.8584       & 0.9216       \\ \bottomrule
\end{tabular}
\end{table*}

The 9 cascade classifiers are tested by the train and test sets, the testing scores are shown in Table ~\ref{table:testing_scores_of_cascade}. For the testing scores of the total train sets, the accuracies are in interval [0.9033, 0.9549], $S$ scores are in interval $[0.8964, 0.9472)$ and $C$ scores are in interval [0.7205,0.7974]) , which are different from the 5-fold cross validation scores when they are trained separately. For example, the accuracy and $C$ score of KNN+RF are 0.9388 and 0.7974, while in separate training, the 5-fold cross validation accuracy and $C$ score of KNN are 0.9867 and 0.9813, the 5-fold cross validation accuracy and $C$ score of RF are 0.8267 and 0.8598. Thus, it is necessary to test the cascade classifiers using the total train sets.

KNN+ANN are considered as the best cascade classifier. The reasons are, on the test sets, the $S$ score of KNN+ANN is 0.7616, which is obviously higher than the other models (their scores are in interval $[0.6555, 0.7135])$. $C$ score of KNN+ANN is 0.8736, indicating the good consistency. On the train sets, the $S$ and $C$ scores of KNN+ANN are 0.911 and 0.7282, which also shows good performance on the train sets compared with other models. 

Table ~\ref{table:testing_scores_of_cascade} shows that the proposed OPAI is effective. By only considering the accuracy, model selection could be problematic. For example, the accuracies of RF+KNN, ANN+ANN and ANN+RF are the same 0.7061, and are higher than those of other models. But their $S$ scores are lower compared with other model combinations, e.g., KNN+ANN, RF+ANN. Lower $S$ scores indicate risky deployment. By considering $S$ and $C$ scores, we can accurately select best model with more angles.

\begin{table*}[]
\caption{ The testing scores of cascade classifiers}
\label{table:testing_scores_of_cascade}
\centering
\begin{tabular}{@{}ccccccccccc@{}}
\toprule
\textbf{Dataset}                                  &                   & \textbf{KNN+KNN} & \textbf{KNN+RF} & \textbf{KNN+ANN} & \textbf{RF+KNN} & \textbf{RF+RF} & \textbf{RF+ANN} & \textbf{ANN+KNN} & \textbf{ANN+RF} & \textbf{ANN+ANN} \\ \midrule
\multirow{3}{*}{\textbf{\begin{tabular}[c]{@{}c@{}}Test \\ (M1-M5, \\ M8-M12)\end{tabular}}}  & \textbf{Accuracy} & 0.9549           & 0.9388          & 0.9321           & 0.9257          & 0.9106         & 0.9033          & 0.9257           & 0.9257          & 0.9033           \\ \cmidrule(l){2-11} 
                                                  & \textbf{$S$}        & 0.9472           & 0.9253          & 0.911            & 0.9316          & 0.9128         & 0.8964          & 0.9316           & 0.9316          & 0.8964           \\ \cmidrule(l){2-11} 
                                                  & \textbf{$C$}        & 0.7595           & 0.7974          & 0.7282           & 0.741           & 0.7913         & 0.7205          & 0.741            & 0.741           & 0.7205           \\ \midrule
\multirow{3}{*}{\textbf{\begin{tabular}[c]{@{}c@{}}Test\\  (M6, M7, \\ M13, M14)\end{tabular}}} & \textbf{Accuracy} & 0.6969           & 0.6921          & 0.6867           & 0.7061          & 0.6677         & 0.6853          & 0.7061           & 0.7061          & 0.6853           \\ \cmidrule(l){2-11} 
                                                  & \textbf{$S$}        & 0.7135           & 0.6983          & 0.7616           & 0.6555          & 0.675          & 0.6915          & 0.6555           & 0.6555          & 0.6915           \\ \cmidrule(l){2-11} 
                                                  & \textbf{$C$}        & 0.7199           & 0.8879          & 0.8736           & 0.9541          & 0.9998         & 0.9946          & 0.9541           & 0.9541          & 0.9946           \\ \bottomrule
\end{tabular}
\end{table*}

\subsection{Comparing and Discussion}

\begin{figure}[]
    \centering
    \includegraphics[width=1\linewidth]{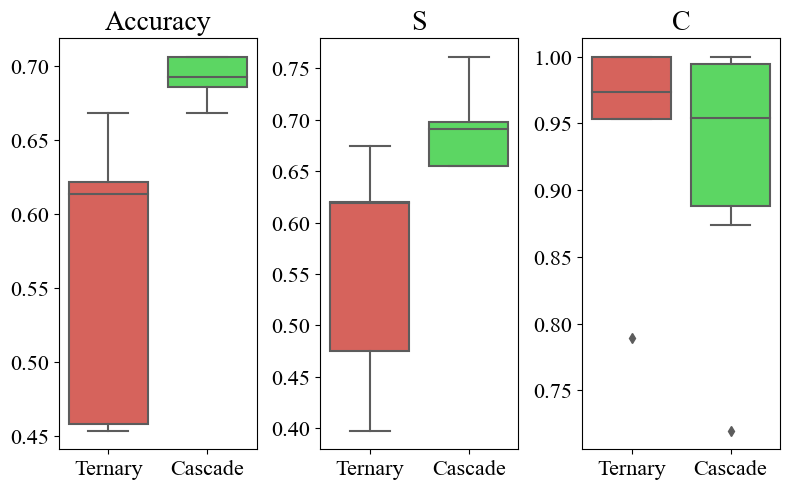}
    \caption{Boxplots of testing scores for ternary classifiers and cascade classifiers}
    \label{fig: boxplot_testing_scores}
\end{figure}

To compare the performance of ternary and cascade classifications on the test sets, we visualized the testing accuracy, S and C scores in Table ~\ref{table:scores_of_ternary_classifiers} and Table ~\ref{table:testing_scores_of_cascade}. Figure ~\ref{fig: boxplot_testing_scores} shows the boxplots of the testing accuracy, $S$ and $C$ scores for the ternary and cascade classifications. We can observe that the median testing accuracy and $S$ scores for the cascade classification are around 0.69, which is obviously higher than the median testing accuracy and $S$ scores for the ternary classification (around 0.61). Compared with ternary classifiers, testing accuracy and $S$ scores of cascade classifiers are higher and less variable. 

The consistency of selected ternary and cascade classifiers are acceptable. Only two classifiers have a $C$ score slightly below 0.85 (See 2 outliers in the right part of Figure ~\ref{fig: boxplot_testing_scores}). In conclusion, the cascade classifiers performs better than the ternary classifiers.

\begin{figure}[]
    \centering
    \includegraphics[width=1\linewidth]{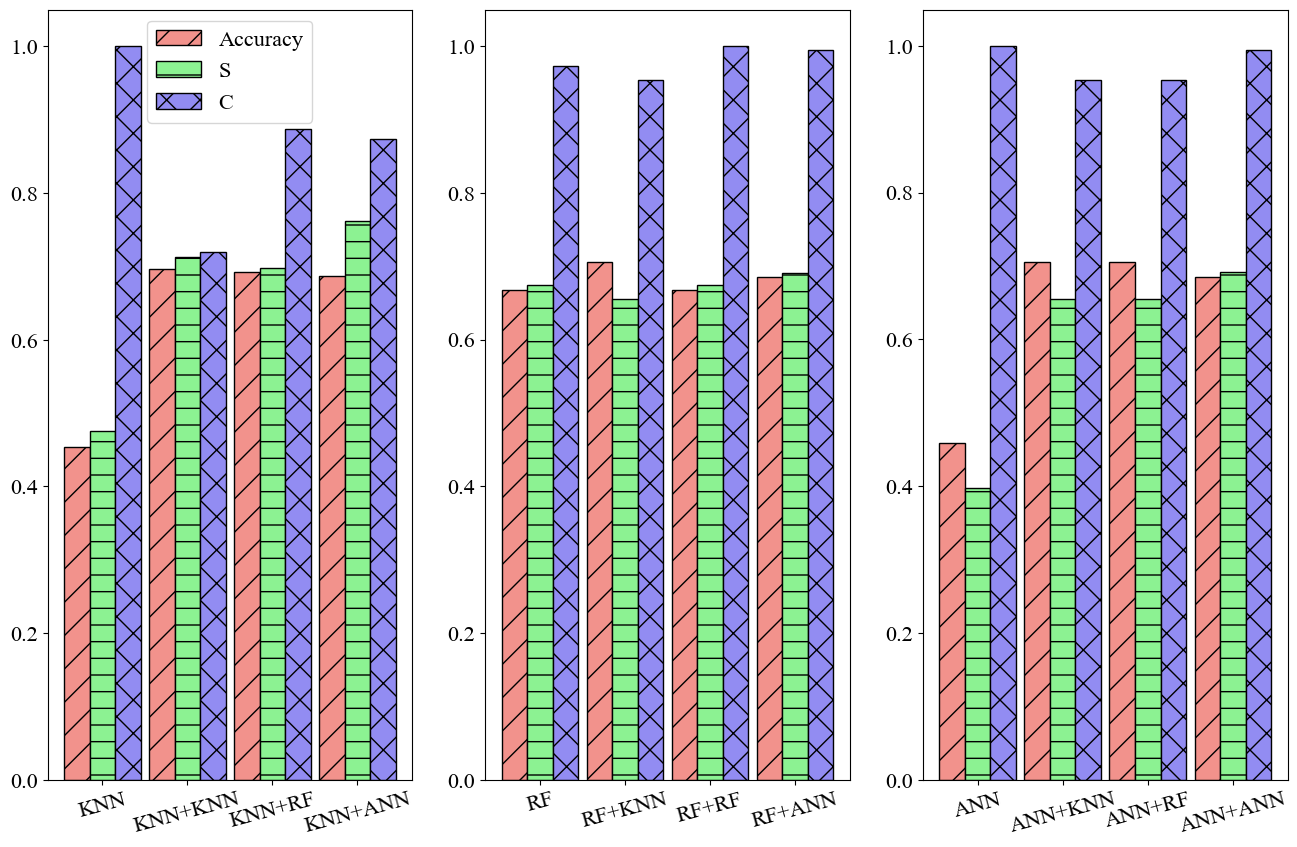}
    \caption{Grouped bar charts of testing scores for ternary classifiers and cascade classifiers}
    \label{fig: bar_charts}
\end{figure}

To further validate whether the cascade classifying process can improve the model’s predictive ability, we compare the testing scores of single ternary classifying algorithm and its N-A classifier combined with other best R-H classifiers. KNN, RF and ANN are selected in the cascade classification because of their good performance. We visualize the ternary and cascade testing scores of KNN, RF and ANN in Table ~\ref{table:scores_of_ternary_classifiers} and Table ~\ref{table:testing_scores_of_cascade} in groups, the grouped bar charts are shown in Figure ~\ref{fig: bar_charts}.

For the KNN and ANN algorithms, the accuracy and $S$ scores of all the combined models are significantly improved over the ternary classifiers, e.g., the ternary classifying testing $S$ scores of KNN and ANN are 0.4753 and 0.3978, the testing scores are not good. But the testing $S$ score of KNN+ANN is 0.7616, which is significantly improved and shows good generalization ability. 

Compared with RF ternary classifier, the improvement of accuracies and $S$ scores for the RF based model combinations is not obvious. At least, it demonstrates that the cascade classification process does not degrade the model prediction performance. In summary, the cascade classification process is effective.

%% file: sections/conclusion.tex
\section{conclusion}\label{Sec:5}
In this paper, we propose a new predictive modeling method and process improvement in the field of data-driven fault prediction. The cascade classification process is proposed to optimize the fault prediction modeling. First, the N-A classifier is used to diagnose whether the equipment is normal and identify early deterioration. Once potential failures are detected, the R-H classifier is activated to further classify the trend of deterioration, i.e., diagnose the risky or high-risk deterioration. By comparing the ternary classifiers with the cascade classifiers, it is verified that the cascade classification process significantly improves the generalization ability of the prediction model. 

At the same time, an innovative design of the OPAI for prediction models is developed. The OPAI can fully consider the time-series distribution characteristics of the predictive modeling data and the online predictive stability of the predictive models. Two evaluation indexes in OPAI can accurately evaluate and select best predictive models. The above innovation is validated on real rotor vibration data of large rotating machines.

The study also has some limitations. Due to data qualities, it is impossible for us to consider the influences of equipment types, potential fault root causes, fault types and locations, sensor measurement and time interval errors. These problems could lead to unstable and inaccurate model predictions. However, we have verified the validity of the cascade classification process by conducting numerous computational experiments and participating in the real-world industrial big data competitions, which is not fully discussed in this paper. Future research direction could consider the different deterioration characteristics and design sub-processes based on these characteristics and fault physics.